\documentclass{jpsj3}
\usepackage{txfonts}
\usepackage{bm}

\usepackage{color}

\title{Solving Graph Coloring Problems Using Feedback-Based Algorithm for Quantum Optimization}

\author{Yuika Kawatomi, Yukina Tatsuta, and Kazue Kudo\thanks{kudo@is.ocha.ac.jp}}
\inst{Department of Computer Science, Ochanomizu University, Bunkyo, Tokyo 112-8610, Japan} 

\abst{The feedback-based algorithm for quantum optimization (FALQON), which is an extension of the quantum approximate optimization algorithm (QAOA), is a promising approach for finding approximate solutions to combinatorial optimization problems.  
The FALQON avoids the optimization of variational parameters by using information derived from iterative measurements.
To solve a constrained combinatorial optimization problem, it is sometimes more efficient to use an XY mixer rather than the conventional X mixer.
We applied the FALQON with the XY mixer to graph coloring and investigated the expectation value of the problem Hamiltonian, success probability of graph coloring, and fair sampling for several small-scale problems.
The FALQON with the XY mixer exhibited better performance than that with the X mixer.
}

\begin{document}
\maketitle

\section{Introduction}
In recent years, solving combinatorial optimization problems using quantum computing has attracted considerable interest. 
Quantum annealing (QA) is an efficient method for finding approximate solutions to these problems~\cite{kadowaki_1998}. 
To solve combinatorial optimization problems using QA, the problem should be formulated as a problem Hamiltonian, which is in the form of an Ising model, such that the ground state of the problem Hamiltonian corresponds to the solution of the problem.

The quantum approximate optimization algorithm (QAOA) and its variants are gate-model algorithms used to obtain approximate solutions to combinatorial optimization problems~\cite{blekos_2024}.
The QAOA uses parametric quantum circuits and requires the optimization of variational parameters.
In contrast, the feedback-based algorithm for quantum optimization (FALQON), an extension of the QAOA, avoids the optimization of variational parameters by utilizing information derived from iterative measurements~\cite{magann_2022a,magann_2022b}.
The computational overhead of FALQON is significantly reduced compared with that of QAOA, although practical deployment on actual hardware incurs additional costs owing to the evaluation of the feedback term.
Several extensions of FALQON have been proposed recently~\cite{arai_2025,abdul_rahman_2025,abdul_rahman_2026}.

Graph coloring is a typical combinatorial optimization problem that requires assigning colors to the nodes of a graph such that no adjacent nodes have the same color.
To solve a graph coloring problem using the QA or QAOA, we typically use one-hot encoding and impose a one-hot constraint that enforces each node to have exactly one color.
The standard approach to implementing one-hot constraints is to add penalty terms to the problem Hamiltonian such that the expectation values of the problem Hamiltonian increase when constraints are violated.
As another approach, we can employ constrained QA, where the driver Hamiltonian is chosen such that the constraints are naturally satisfied without penalty terms~\cite{hen_2016a, hen_2016b, kudo_2018, kudo_2020}.
In the context of graph coloring, the driver Hamiltonian is represented as a variant of the XY model instead of the transverse X driver.
Similar QAOA approaches have been proposed, including the QAOA using the XY mixer~\cite{hadfield_2019, wang_2020} and the fermionic QAOA~\cite{yoshida_2023}.

In this study, we applied the FALQON with the XY mixer to solve graph coloring problems. 
Although a recent study on portfolio optimization using FALQON variants with the XY mixer demonstrated some advantages of the XY mixer~\cite{prado_2026}, the feedback Hamiltonian was not explicitly presented. 
In this study, we derived the feedback Hamiltonian for the XY mixer for graph coloring.
The noiseless simulation results of the X and XY mixers demonstrated some advantages of the XY mixer in the success probability of correct coloring. 
We also compared the performances of FALQON and QAOA results. 
Furthermore, we explored fair sampling, specifically whether the degenerate ground states were equally sampled.

The remainder of this paper is organized as follows. 
Section~\ref{sec:model_method} describes the FALQON and QAOA models for graph coloring and the numerical procedures used to obtain the results. 
Section~\ref{sec:result} presents the results, that is, the expectation values of the problem Hamiltonian and the success probability, obtained using the FALQON and QAOA with the XY and X mixers. 
The fair sampling results obtained using FALQON are also presented. 
Finally, the conclusions are presented in Sect. ~\ref{sec:conclusion}.

\section{Models and methods}
\label{sec:model_method}

\subsection{Feedback-based algorithm for quantum optimization}
\label{ssec:FALQON}

FALQON is based on the quantum dynamics governed by 
\begin{align}
i\frac{d}{dt} \left|\psi(t)\right\rangle 
= [H_{\rm p}+H_{\rm m}\beta(t)] \left|\psi(t)\right\rangle,
\end{align}
where $\left|\psi(t)\right\rangle$ is the state vector, $H_{\rm p}$ and $H_{\rm m}$ are the problem and mixer Hamiltonians, respectively, and $\beta(t)$ is a scalar, time-dependent function.
Here, we set $\hbar=1$.
To minimize the expectation value $\langle H_{\rm p}\rangle = \langle\psi(t)|H_{\rm p}|\psi(t)\rangle$, $\beta(t)$ is designed to satisfy the following condition: 
\begin{align}
\frac{d}{dt}\langle\psi(t)|H_{\rm p}|\psi(t)\rangle\le 0, \quad
\forall t\ge 0.
\label{eq:dec_cond}
\end{align}
The time evolution is approximated by Trotterization, which is performed by applying
$U_{\rm p}=e^{-iH_{\rm p}\Delta t}$ and 
$U_{\rm m}(\beta_k)=e^{-i\beta_k H_{\rm m}\Delta t}$ alternately for $k=1,2,\ldots, L$, 
where $L$ is the number of layers, and the time duration $\Delta t$ is small.
Here, $\beta_k$ is determined by the feedback law given by
\begin{align}
\beta_{k+1} = -A_k=-\langle\psi_k|H_{\rm f}|\psi_k\rangle,
\end{align}
where $H_{\rm f}=i[H_{\rm m},H_{\rm p}]$ is the feedback Hamiltonian, and $|\psi_k\rangle=U_{\rm m}(\beta_k)U_{\rm p}\cdots U_{\rm m}(\beta_1)U_{\rm p} \left|\psi_0\right\rangle$, where $|\psi_0\rangle$ is the initial state.

In the case of QAOA, the state at the end of the quantum circuit is expressed as $|\psi(\bm{\gamma},\bm{\beta})\rangle = U_{\rm m}(\beta_L)U_{\rm p}(\gamma_L) \cdots U_{\rm m}(\beta_1)U_{\rm p}(\gamma_1)|\psi_0\rangle$, where $U_{\rm p}(\gamma_j)=e^{-i\gamma_j H_{\rm p}}$ and $U_{\rm m}(\beta_j)=e^{-i\beta_j H_{\rm m}}$ for $j=1, 2, \ldots, L$.
To minimize $\langle H_{\rm p}\rangle$, expressed as $\langle\psi(\bm{\gamma},\bm{\beta})|H_{\rm p}|\psi(\bm{\gamma},\bm{\beta})\rangle$, the parameters $\bm{\gamma}$ and $\bm{\beta}$ are optimized simultaneously using a classical processor, whereas FALQON employs a feedback mechanism based on measurements for each layer, thus eliminating the need for classical optimizations.

\subsection{Models for graph coloring}
\label{ssec:model}

First, we express the objective function to minimize, using binary variables.
The term to avoid coloring adjacent nodes in the same color is described as
\begin{align}
E_{\rm adj} = \sum_{(v,w)\in \mathcal{E}}\sum_{c=1}^K q_{v,c}q_{w,c}.
\end{align}
Here, $(v,w)$ denotes the edge connecting nodes $v$ and $w$, $\mathcal{E}$ is the set of edges, and $q_{v,c}=1$ if node $v$ has color $c$; otherwise, $q_{v,c}=0$.
The term to impose one-hot constraints, i.e, each node has just one color, is described as
\begin{align}
E_{\rm one} = \sum_{v\in \mathcal{V}}\left(\sum_{c=1}^K q_{v,c}-1\right)^2,
\end{align}
where $\mathcal{V}$ denotes the node set. 
When the graph is colored correctly, $E_{\rm adj}=E_{\rm one}=0$.
By replacing $q_{v,c}$ with $\frac12(1-Z_{v,c})$, we obtain the Hamiltonians corresponding to $E_{\rm adj}$ and $E_{\rm one}$ as follows:
\begin{align}
H_{\rm adj} &= \frac14 \sum_{(v,w)\in \mathcal{E}}\sum_{c=1}^K 
(1-Z_{v,c})(1-Z_{w,c}),
\\
H_{\rm one} &=  \sum_{v\in \mathcal{V}}\left( \frac12\sum_{c=1}^K (1-Z_{v,c})-1\right)^2.
\end{align}
Here, $X_{v,c}$, $Y_{v,c}$ and $Z_{v,c}$ are the Pauli $Z$ operators acting on the qubit corresponding to $q_{v,c}$.

When we use the X mixer, the problem and mixer Hamiltonians are given as
\begin{align}
H_{\rm p}^{X} &= H_{\rm adj} + \lambda H_{\rm one},
\\
H_{\rm m}^{X} &= \sum_{v\in \mathcal{V}}\sum_{c=1}^K X_{v,c},
\end{align}
respectively, where $\lambda$ is a positive parameter corresponding to the strength of the constraint.
In contrast, when we use the XY mixer, we can eliminate the one-hot constraint term in the problem Hamiltonian.
Then, we define the problem and mixer Hamiltonians as
\begin{align}
H_{\rm p}^{XY} &= H_{\rm adj},
\\
H_{\rm m}^{XY} &= \frac12\sum_{v\in \mathcal{V}}\sum_{1\le a<b\le K} (X_{v,a}X_{v,b}+Y_{v,a}Y_{v,b}),
\end{align}
respectively.
The summation regarding the color index in $H_{\rm m}^{XY}$ is taken for all combinations where $a < b$.

The feedback Hamiltonian in FALQON for the X mixer is written as
\begin{align}
\label{eq:H_f_X}
H_{\rm f}^X &\equiv i[H_{\rm m}^X, H_{\rm p}^X]
\nonumber \\
&=\frac12\sum_{(v,w)\in \mathcal{E}}\sum_{c=1}^K (Y_{v,c}Z_{w,c} + Z_{v,c}Y_{w,c} - Y_{v,c} - Y_{w,c})
\nonumber\\
&\quad
+\lambda\sum_{v\in V}\left\{
\sum_{1\le a<b \le K} (Y_{v,a}Z_{v,b} + Z_{v,a}Y_{v,b})
- (K-2)\sum_{c=1}^K Y_{v,c}
\right\},
\end{align}
and that for the XY mixer is derived as
\begin{align}
\label{eq:H_f_XY}
H_{\rm f}^{XY} &\equiv i[H_{\rm m}^{XY}, H_{\rm p}^{XY}]
\nonumber\\
&= -\frac14\sum_{(v,w)\in \mathcal{E}}\sum_{1\le a<b\le K}\left[
(X_{v,a}Y_{v,b}-Y_{v,a}X_{v,b})(Z_{w,a}-Z_{w,b})
+(X_{w,a}Y_{w,b}-Y_{w,a}X_{w,b})(Z_{v,a}-Z_{v,b})
\right].
\end{align}
The details of the derivation are provided in the Appendix.

\subsection{Numerical procedures}
\label{ssec:numerical}

The initial state is usually prepared as a uniform superposition when using the X mixer.
In this study, we followed this convention, and the initial state was given as
\begin{align}
|\psi_0^X\rangle = |+\rangle^{\otimes NK},
\end{align}
where $|+\rangle=(|0\rangle+|1\rangle)/\sqrt{2}$, and $N$ is the number of nodes.
However, when the XY mixer is used, the initial state must satisfy the one-hot constraints.
Considering the $W$ state of $K$ qubits for each node,
\begin{align}
\left| W_K\right\rangle =\frac{1}{\sqrt{K}}\sum_{c=1}^K
|0\cdots \overset{c}{1} \cdots 0\rangle,
\end{align}
where $|0\cdots \overset{c}{1} \cdots 0\rangle$ denotes the state where the $c$th qubit is $|1\rangle$ and the other qubits are $|0\rangle$,
we prepared the initial state for the XY mixer as
\begin{align}
|\psi_0^{XY}\rangle = |W_K\rangle^{\otimes N}.
\end{align}
This state, which is a superposition of states in which each node has only one color, satisfies the one-hot constraints.

We conducted computations employing the FALQON with the X mixer (X-FALQON) and XY mixer (XY-FALQON) for up to $L=800$ layers.
The initial parameter of $\beta$ was set to $\beta_0=0$.
The time step was set to $\Delta t=0.025$, which was sufficiently small to satisfy Eq.~\eqref{eq:dec_cond}. 
This was verified using preliminary simulations.
Although $\Delta t$ should be tuned for efficient computation, we fixed its value in this study. 

For comparison, we conducted computations using the QAOA with the X mixer (X-QAOA) and XY mixer (XY-QAOA) for $L=2,\, 4,\, 6,\, 8,\, 10$.
The parameters $\beta_1,\ldots,\beta_L$ and $\gamma_1,\dots,\gamma_L$ were optimized classically using the constrained optimization by linear approximation (COBYLA).
For $L=2$, the initial parameter values were chosen uniformly randomly from $0\le\gamma_k\le 4\pi$ and $0\le\beta_k\le\pi$, where $k=1,\ldots, L$, for the X-QAOA, and from $0\le\gamma_k\le 2\pi$ and $0\le\beta_k\le 2\pi$ for the XY-QAOA.
For $L\ge 4$, we initialized the parameters using linear interpolation with the optimized parameter values for $L-2$.
Suppose we initialize the parameters for $L_{\rm new}$ layers.
Consider the optimized parameters as $\gamma^*_1,\dots,\gamma^*_{L_{\rm old}}$ for $L_{\rm old}=L_{\rm new}-2$.
For $k =1,\ldots, L_{\rm new}-1$, we set the initial values as
\begin{align}
\gamma_k = (1-t_k)\gamma^*_j + t_k\gamma^*_{j+1},\quad
t_k=r_k-j,
\end{align}
where $j=\lfloor r_k \rfloor$ with $\lfloor\cdot\rfloor$ being the floor function, and
\begin{align}
r_k = 1 + (k-1)\frac{L_{\rm old}-1}{L_{\rm new}-1}.
\end{align}
For $k=L_{\rm new}$, $\gamma_{L_{\rm new}}=\gamma^*_{L_{\rm old}}$.
Additionally, a Gaussian random number with a mean of $0$ and a variance of $0.0025$ was added to each parameter value.
The initial parameter values of $\beta_1,\ldots,\beta_{L_{\rm new}}$ were given in a similar manner.
This parameter initialization method is more efficient than random initialization and is useful for shortening the computation time in QAOA.

For both FALQON and QAOA, the coefficient of the one-hot constraint term in the X mixer was fixed to $\lambda=0.5$, although $\lambda$ should be tuned to obtain a better solution. 
We executed FALQON and QAOA using noiseless statevector simulations with Qiskit \cite{qiskit2024}.
Although parameter $\beta_k$ at each layer $k$ in FALQON is computed sequentially in a deterministic manner, parameters $\bm{\gamma}$ and $\bm{\beta}$ in QAOA depend on their initial values. Thus, in the QAOA, we performed 20 executions for each case of $L=2,\, 4,\, 6,\, 8,\, 10$.

\section{Results}
\label{sec:result}

\begin{figure}
    \centering
    \includegraphics[width=0.5\linewidth]{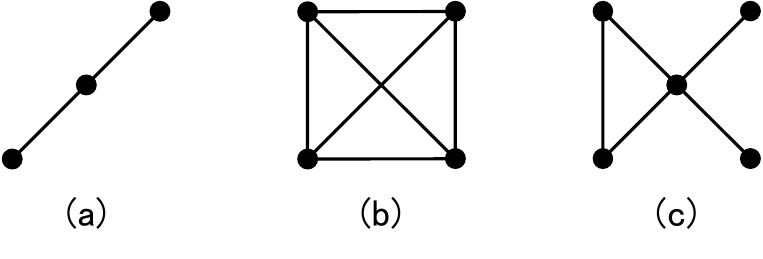}
    \caption{Graphs to be colored. 
    (a) Graph consisting of three nodes and two edges to be colored using up to three colors, although it can be colored using only two colors. 
    (b) Graph consisting of four nodes and six edges to be colored using four colors.
    (c) Graph consisting of five nodes and five edges to be colored in three colors.}
    \label{fig:graphs}
\end{figure}

We solved three graph coloring problems, the graphs of which are shown in Fig.~\ref{fig:graphs}.
The easiest of the three is graph (a).
Although graph (a) can be colored using only two colors, we considered coloring it with up to three colors.
Thus, the ground state of this problem is highly degenerate.
The hardest of the three is graph (b), where the fully connected four-node graph is colored in four colors.
Graph (c), which has five nodes and five edges, is colored with three colors.

\subsection{Expectation values of the problem Hamiltonians}
\label{ssec:expectation}

\begin{figure*}
    \centering
    \includegraphics[width=1.0\linewidth]{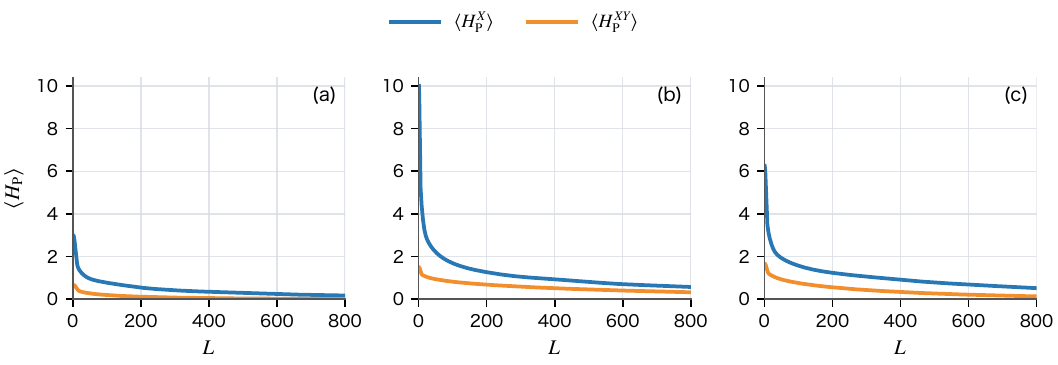}
    \caption{Expectation values of the problem Hamiltonians for FALQON as functions of the number of layers.
    The blue and orange curves represent the results of X-FALQON and XY-FALQON, respectively.
    The problems solved in (a), (b) and (c) correspond to Figs.~\ref{fig:graphs}(a), \ref{fig:graphs}(b) and \ref{fig:graphs}(c), respectively.}
    \label{fig:Hp_falqon}
\end{figure*}

The problem Hamiltonians corresponding to the X and XY mixers are different from each other.
However, their expectation values, $\langle H_{\rm p}^{X}\rangle$ and $\langle H_{\rm p}^{XY}\rangle$, are zero in the ground state.
As shown in Fig.~\ref{fig:Hp_falqon}, $\langle H_{\rm p}^{X}\rangle$ and $\langle H_{\rm p}^{XY}\rangle$ for X-FALQON and XY-FALQON decay with increasing number of layers for the three problems.
We observe an initial rapid decay followed by a gradual decrease in all cases, whereas the values at $L=800$ vary depending on the difficulty level of the problems and mixers.

\begin{figure}
    \centering
    \includegraphics[width=0.5\linewidth]{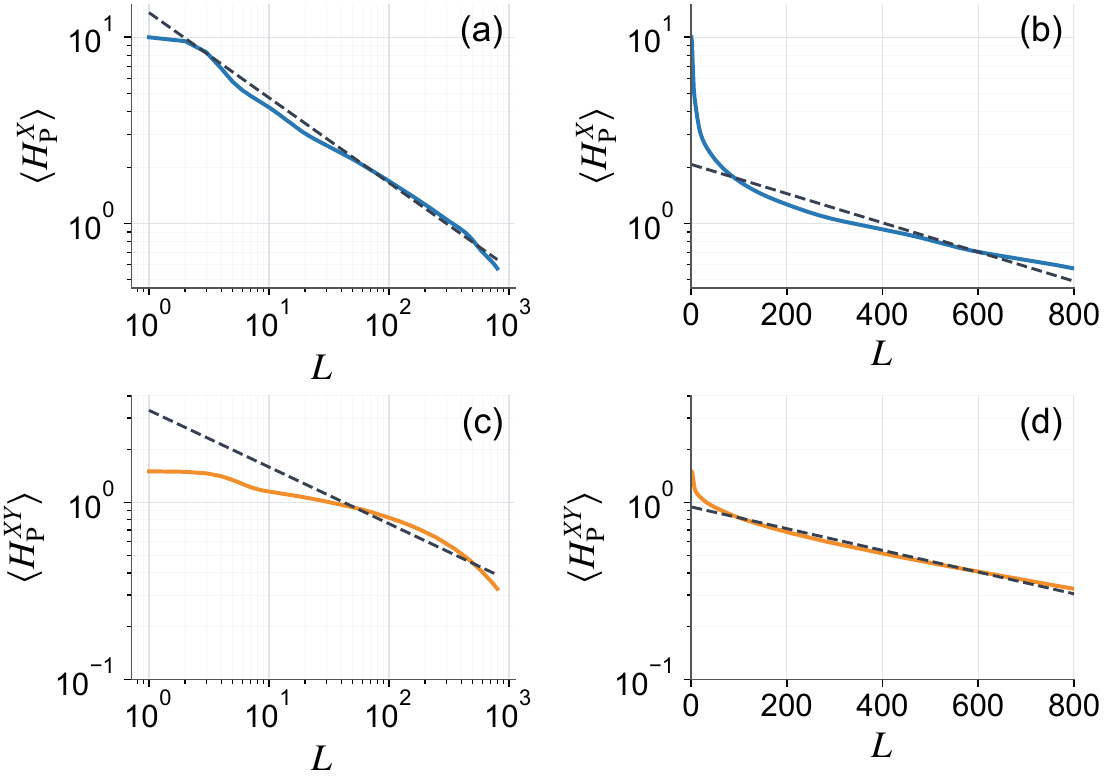}
    \caption{Log-log and semi-log plots of Fig.~\ref{fig:Hp_falqon}(b) and their regression curves. (a) and (b) plot the X-FALQON result, while (c) and (d) plot the XY-FALQON result.}
    \label{fig:Hp_falqon_log}
\end{figure}

The decay rate of the expectation value also varied depending on the type of mixer.
Figure~\ref{fig:Hp_falqon_log} shows the log-log and semi-log plots of Fig.~\ref{fig:Hp_falqon}(b).
The dashed lines represent the regression curves.
The X-FALQON result fits the dashed line of the log-log plot, whereas the XY-FALQON result fits that of the semi-log plot.
In other words, $\langle H_{\rm p}^{X}\rangle$  and $\langle H_{\rm p}^{XY}\rangle$ seem to decay polynomially and exponentially, respectively.

The tendency of decay in $\langle H_{\rm p}^{X}\rangle$ and $\langle H_{\rm p}^{XY}\rangle$ is common to all three problems.
We examined the coefficient of determination, which is defined by
\begin{align}
R^2 = 1 - \frac{\sum_{l=1}^L(y_l-f(l))^2}{\sum_{l=1}^L(y_l-\bar{y})^2},
\end{align}
where $\bar{y}=\frac{1}{L}\sum_{l=1}^L y_l$.
In this section, $y_l$ is $\langle H_{\rm p}\rangle$ of the $l$th layer, and $f(x)$ is an exponential or power function.
The higher $R^2$, the better the regression, and $R^2=1$ in the best case.
As shown in Table~\ref{tab:R2}, for the X mixer, the $R^2$ value for the power function was higher than that for the exponential function. 
In contrast, for the XY mixer, the $R^2$ value for the exponential function was higher than that for the power function.
In other words, $\langle H_{\rm p}^X\rangle$ decreases according to a power law, and $\langle H_{\rm p}^{XY}\rangle$ decays exponentially, at least for $L\le 800$.
Because the decay in the expectation value did not converge at $L=800$, the above result does not necessarily hold for a much larger $L$.

\begin{table}
\caption{Coefficients of determination $R^2$. 
Data in Fig.~\ref{fig:Hp_falqon} were fitted to power and exponential functions.}
\label{tab:R2}
\centering
    \begin{tabular}{c|c|c|c}
    \hline
      Mixer & Problem & $R^2$ (power) & $R^2$ (exp.) \\
      \hline
      X & (a) & 0.945 & 0.924\\
      X & (b) & 0.985 & 0.851\\
      X & (c) & 0.955 & 0.905\\
      \hline
     XY & (a) & 0.863 & 0.983\\
     XY & (b) & 0.897 & 0.967\\
     XY & (c) & 0.857 & 0.986\\
     \hline
    \end{tabular}
\end{table}

\begin{figure*}
    \centering
    \includegraphics[width=1.0\linewidth]{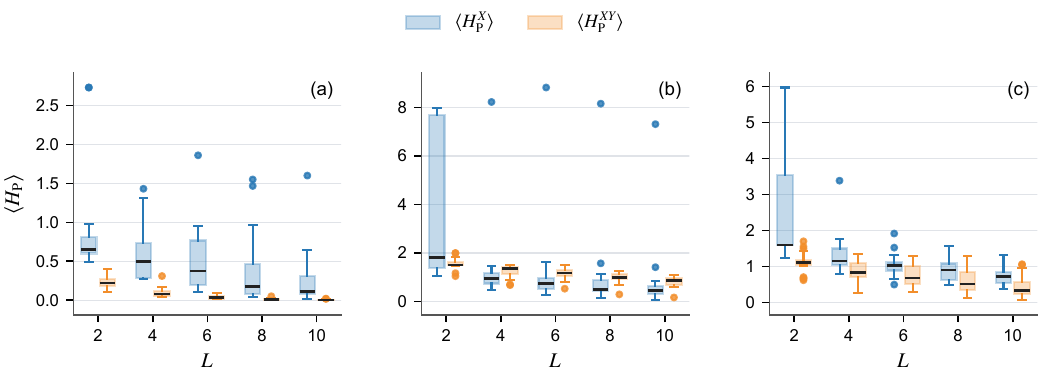}
    \caption{Box-and-whisker plots of expectation values of the problem Hamiltonians for the QAOA with $L=2,\, 4,\, 6,\, 8,\, 10$.
    The box plots indicate the median (horizontal line), 25th and 75th percentiles (box), and whiskers extend to the furthest data points within $1.5$ times the interquartile range (IQR) from the quartiles; data points beyond this range are plotted individually.
    The blue and orange boxes represent the results of X-QAOA and XY-QAOA, respectively.
    The problems solved in (a), (b) and (c) correspond to Figs.~\ref{fig:graphs}(a), \ref{fig:graphs}(b) and \ref{fig:graphs}(c), respectively.}
    \label{fig:Hp_qaoa}
\end{figure*}

Figure~\ref{fig:Hp_qaoa} illustrates $\langle H_{\rm p}^X\rangle$ and $\langle H_{\rm p}^{XY}\rangle$ for the X-QAOA and XY-QAOA.
Although the expectation values varied depending on the initial parameters, the median value decreased as the number of layers increased.

\subsection{Success probability}
\label{ssec:success}

The success probability is the probability of measuring the degenerate ground state, which corresponds to the probability of obtaining a solution to the graph coloring problem considered:
\begin{align}
P_{\rm suc} = \sum_{q\in S} \left|\langle q|\psi\rangle\right|^2,
\end{align}
where $S$ denotes the set of states corresponding to the correct coloring.

\begin{table}
    \caption{Success probabilities for the FALQON and QAOA with the X and XY mixers.
    The problems solved in (a), (b) and (c) correspond to Figs.~\ref{fig:graphs}(a), \ref{fig:graphs}(b) and \ref{fig:graphs}(c), respectively.
    Each result for QAOA is the median of 20 results for $L=10$, and that for FALQON is the value at $L=800$.}
    \label{tab:P}
    \centering
    \begin{tabular}{c|cc|cc}
    \hline
    Problem & X-FALQON & XY-FALQON & X-QAOA & XY-QAOA \\
    \hline
    (a) & 72.0\% & 98.9\% & 86.9\% & 99.6\% \\
    (b) & 40.7\% & 69.3\% & 49.9\% & 38.2\% \\
    (c) & 39.7\% & 87.7\% & 20.1\% & 73.8\% \\
    \hline
    \end{tabular}
\end{table}

Table~\ref{tab:P} shows the success probabilities for the FALQON and QAOA with the X and XY mixers.
The results for FALQON were evaluated at $L=800$, whereas for QAOA, the median of 20 results for $L=10$ is presented.
In FALQON, the results obtained with the XY-mixer outperformed those obtained with the X-mixer across all three problems, indicating the advantages of using the XY-mixer to solve graph coloring problems.
For QAOA, the X-mixer result surpassed the XY-mixer result in problem (b), although the XY-mixer results were better than those of the X-mixer in problems (a) and (c).
This result may reflect the following facts: The QAOA results heavily depend on parameter optimization, and the constraint parameter $\lambda$, which should be tuned according to the problem, was fixed in this study.
XY-FALQON shows the highest success probability, except for problem (a), which is the easiest of the three problems.
Because the FALQON algorithm did not converge at $L=800$, the success probabilities using FALQON are expected to be higher for larger $L$.

\subsection{Fair sampling}
\label{ssec:fair}

In our formulation, there are several solutions for each graph coloring problem.
In other words, the problem Hamiltonian has degenerate ground states.
Ground states are not always fairly sampled (i.e., sampled with equal probabilities) in FALQON or QAOA.
It is known that fair sampling in QAOA depends on the type of mixer Hamiltonian~\cite{golden_2022}.
Here, we examined the performance of X-FALQON and XY-FALQON in terms of fair sampling.

To quantify fair sampling, we introduce the root mean square of relative error defined by
\begin{align}
R_{\rm err} = \sqrt{\frac{1}{N_S}\sum_{q\in S}\left(
\frac{p_q-p^*}{p^*}\right)^2},
\label{eq:R_err}
\end{align}
where $S$ and $N_S$ denote the set of states corresponding to the correct coloring and the number of elements in the set, respectively.
Here, $p_q=\left|\langle q|\psi\rangle\right|^2$ is the probability of measuring state $|q\rangle$, and $p^*=1/N_S$ is the expected probability for the fair sampling.

\begin{figure*}
    \centering
    \includegraphics[width=1.0\linewidth]{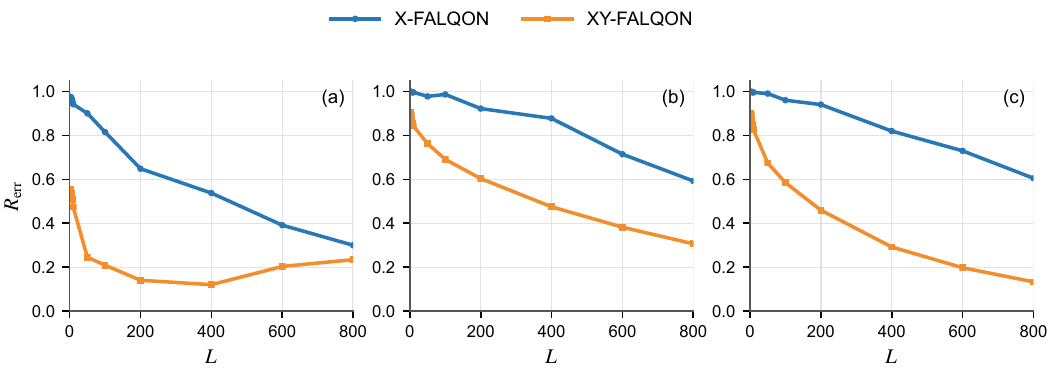}
    \caption{Root mean square of relative error defined by Eq.~\eqref{eq:R_err} as a function of the number of layers, $L$. 
    The blue and orange curves represent the results of X-FALQON and XY-FALQON, respectively. 
    The problems solved in (a), (b) and (c) correspond to Figs.~\ref{fig:graphs}(a), (b) and (c), respectively.}
    \label{fig:rmspe}
\end{figure*}

Figure~\ref{fig:rmspe} demonstrates $R_{\rm err}$ of X-FALQON and XY-FALQON.
For all problems, $R_{\rm err}$ of XY-FALQON is smaller than that of X-FALQON, which indicates the advantages of XY-FALQON in fair sampling.
Figures~\ref{fig:rmspe}(b) and \ref{fig:rmspe}(c) illustrate that $R_{\rm err}$ decays almost monotonically with increasing $L$ for both X-FALQON and XY-FALQON.
However, in Fig.~\ref{fig:rmspe}(a), $R_{\rm err}$ of XY-FALQON shows a small increase after $L=400$, although that of X-FALQON decreases monotonically.
Even in the region where $R_{\rm err}$ increases, $\langle H_{\rm p}^{XY}\rangle$ decreases, as shown in Fig.~\ref{fig:Hp_falqon}(a).
This result implies that a high success probability does not always result in a fair sampling. 
In the case of (a), the ground states are highly degenerate, resulting in a small $p^*$.
Then, a small fluctuation or weak localization of probability amplitudes can cause the rise of $R_{\rm err}$ in Fig.~\ref{fig:rmspe}(a).

\section{Conclusions}
\label{sec:conclusion}

We applied the XY mixer in FALQON to solve graph coloring problems.
Unlike the X mixer, the XY mixer can avoid the constraint term in the problem Hamiltonian; thus, there is no need to tune the constraint parameter. 
Moreover, the FALQON does not require the optimization of variational parameters, unlike the QAOA.
Thus, FALQON with the XY mixer has advantages in solving constrained combinatorial problems, such as graph coloring.

Using the XY mixer in the FALQON results in higher success probabilities compared to the use of the X mixer.
The expectation value of the problem Hamiltonian exhibited approximately exponential decay with an increasing number of layers.
The XY mixer is also advantageous for fair sampling. 
These results indicate that FALQON using the XY mixer is particularly advantageous for problems with one-hot constraints.

\begin{acknowledgment}
This work was partially supported by JSPS KAKENHI Grant Number JP25H01522.
\end{acknowledgment}

\appendix
\section{\label{sec:appendix}}

This section provides the derivation of the feedback Hamiltonians in FALQON: Eqs.~\eqref{eq:H_f_X} and \eqref{eq:H_f_XY}.
The feedback Hamiltonian for the X mixer is written as follows:
\begin{align}
\label{eqA:H_f_X}
H_{\rm f}^X &= i[H_{\rm m}^X, H_{\rm p}^X]
\nonumber\\
&= i\left[ \sum_{v\in\mathcal{V}} \sum_{c=1}^K X_{v,c}, \; 
\frac14\sum_{(v,w)\in\mathcal{E}}\sum_{c=1}^K
(Z_{v,c}Z_{w,c} - Z_{v,c} - Z_{w,c}) \right]
\nonumber\\
&\quad
+i\left[ \sum_{v\in\mathcal{V}} \sum_{c=1}^K X_{v,c}, \; 
\frac{\lambda}2\sum_{v\in\mathcal{V}}\left\{
\sum_{1\le a<b\le K}Z_{v,a}Z_{v,b} - (K-2)\sum_{c=1}^K Z_{v,c} \right\}\right].
\end{align}
Using the following relation,
\begin{align}
[X_i, Z_j] &= -2iY_i\delta_{ij},
\label{eqA:XZ}
\end{align}
where $\delta_{ij}$ is the Kronecker delta, we obtain the following equations:
\begin{align}
\left[\sum_v\sum_c X_{v,c},\; \sum_{(v,w)}\sum_c Z_{v,c}Z_{w,c}\right]
&= -2i\sum_{(v,w)}\sum_c (Y_{v,c}Z_{w,c} + Z_{v,c}Y_{w,c}),
\label{eqA:X1}\\
\left[\sum_v\sum_c X_{v,c},\; \sum_{(v,w)}\sum_c Z_{v,c}\right]
&= -2i\sum_{(v,w)}\sum_c Y_{v,c},
\label{eqA:X2}\\
\left[\sum_v\sum_c X_{v,c},\; \sum_v\sum_{a<b} Z_{v,a}Z_{v,b}\right]
&= -2i\sum_v\sum_{a<b} (Y_{v,a}Z_{v,b} + Z_{v,a}Y_{v,b}).
\label{eqA:X3}
\end{align}
Combining Eqs. \eqref{eqA:H_f_X}, \eqref{eqA:X1}, \eqref{eqA:X2} and \eqref{eqA:X3} leads to the feedback Hamiltonian for the X mixer:
\begin{align}
H_{\rm f}^X 
&=\frac12\sum_{(v,w)\in \mathcal{E}}\sum_{c=1}^K (Y_{v,c}Z_{w,c} + Z_{v,c}Y_{w,c} - Y_{v,c} - Y_{w,c})
\nonumber\\
&\quad
+\lambda\sum_{v\in V}\left\{
\sum_{1\le a<b \le K} (Y_{v,a}Z_{v,b} + Z_{v,a}Y_{v,b})
- (K-2)\sum_{c=1}^K Y_{v,c}
\right\}.
\end{align}

The feedback Hamiltonian in FALQON for the XY mixer is written as follows:
\begin{align}
\label{eqA:H_f_XY}
H_{\rm f}^{XY} &= i[H_{\rm m}^{XY}, H_{\rm p}^{XY}]
\nonumber\\
&= i\left[ \frac12\sum_{v\in\mathcal{V}}\sum_{1\le a<b\le K} h_v^{ab}, \; 
\frac14\sum_{(v,w)\in\mathcal{E}}\sum_{c=1}^K
(Z_{v,c}Z_{w,c} - Z_{v,c} - Z_{w,c}) \right],
\end{align}
where we defined $h_v^{ab}\equiv(X_{v,a}X_{v,b}+Y_{v,a}Y_{v,b})$.
Using Eq.~\eqref{eqA:XZ} and the following relation,
\begin{align}
[Y_i, Z_j] &= 2iX_i\delta_{ij},
\label{eqA:YZ}
\end{align}
we obtain the following equations:
\begin{align}
\left[\sum_{v} \sum_{a<b} h_v^{ab},\; 
\sum_{(v,w)}\sum_c Z_{v,c}Z_{w,c}\right]
&= 2i\sum_{(v,w)}\sum_{a<b} \left\{ 
B_v^{ab}(Z_{w,a} - Z_{w,b}) + (Z_{v,a} - Z_{v,b})B_w^{ab}\right\},
\label{eqA:X5}\\    
\left[\sum_{v} \sum_{a<b} h_v^{ab},\; 
\sum_{(v,w)}\sum_c (Z_{v,c} + Z_{w,c})\right]
&= 0,
\label{eqA:X6}
\end{align}
where we defined $B_v^{ab}\equiv X_{v,a}Y_{v,b}-Y_{v,a}X_{v,b}$.
Combining Eqs. \eqref{eqA:H_f_XY}, \eqref{eqA:X5} and \eqref{eqA:X6} leads to the feedback Hamiltonian for the XY mixer:
\begin{align}
H_{\rm f}^{XY} &= -\frac14\sum_{(v,w)\in \mathcal{E}}\sum_{1\le a<b\le K}\left[
(X_{v,a}Y_{v,b}-Y_{v,a}X_{v,b})(Z_{w,a}-Z_{w,b})
+(X_{w,a}Y_{w,b}-Y_{w,a}X_{w,b})(Z_{v,a}-Z_{v,b})
\right],    
\end{align}
where we used $[X_{v,a}Y_{v,b},\; Z_{w,a}]=[Y_{v,a}X_{v,b},\; Z_{w,a}]=0$ because $v\neq w$ in the graph coloring setup.


\begin{thebibliography}{10}

\bibitem{kadowaki_1998}
T.~Kadowaki and H.~Nishimori, Phys. Rev. E {\bfseries 58},  5355 (1998).

\bibitem{blekos_2024}
K.~Blekos, D.~Brand, A.~Ceschini, C.-H. Chou, R.-H. Li, K.~Pandya, and
  A.~Summer, Phys. Rep. {\bfseries 1068},  1 (2024).

\bibitem{magann_2022a}
A.~B. Magann, K.~M. Rudinger, M.~D. Grace, and M.~Sarovar, Phys. Rev. Lett.
  {\bfseries 129},  250502 (2022).

\bibitem{magann_2022b}
A.~B. Magann, K.~M. Rudinger, M.~D. Grace, and M.~Sarovar, Phys. Rev. A
  {\bfseries 106},  062414 (2022).

\bibitem{arai_2025}
D.~Arai, K.~N. Okada, Y.~Nakano, K.~Mitarai, and K.~Fujii, Phys. Rev. Research
  {\bfseries 7},  013035 (2025).

\bibitem{abdul_rahman_2025}
S.~Abdul~Rahman, {\"O}.~Karabacak, and R.~Wisniewski, In R.~Wyrzykowski,
  J.~Dongarra, E.~Deelman, and K.~Karczewski (eds), {\itshape Parallel
  {Processing} and {Applied} {Mathematics}}, 2025, pp. 277--289.

\bibitem{abdul_rahman_2026}
S.~Abdul~Rahman, {\"O}.~Karabacak, and R.~Wisniewski, Future Gener. Comput.
  Syst. {\bfseries 174},  107979 (2026).

\bibitem{hen_2016a}
I.~Hen and F.~M. Spedalieri, Phys. Rev. Applied {\bfseries 5},  034007 (2016).

\bibitem{hen_2016b}
I.~Hen and M.~S. Sarandy, Phys. Rev. A {\bfseries 93},  062312 (2016).

\bibitem{kudo_2018}
K.~Kudo, Phys. Rev. A {\bfseries 98},  022301 (2018).

\bibitem{kudo_2020}
K.~Kudo, J. Phys. Soc. Jpn. {\bfseries 89},  064001 (2020).

\bibitem{hadfield_2019}
S.~Hadfield, Z.~Wang, B.~O’Gorman, E.~G. Rieffel, D.~Venturelli, and
  R.~Biswas, Algorithms {\bfseries 12},  34 (2019).

\bibitem{wang_2020}
Z.~Wang, N.~C. Rubin, J.~M. Dominy, and E.~G. Rieffel, Phys. Rev. A {\bfseries
  101},  012320 (2020).

\bibitem{yoshida_2023}
T.~Yoshioka, K.~Sasada, Y.~Nakano, and K.~Fujii, Phys. Rev. Res. {\bfseries 5},
   023071 (2023).

\bibitem{prado_2026}
P.~M. Prado, R.~S.~d. Carmo, L.~A.~M. Rattighieri, L.~G.~E. Arruda, G.~S.
  Franco, M.~C. de~Oliveira, and F.~F. Fanchini, Quantum Inf. Process.
  {\bfseries 25},  275 (2026).

\bibitem{qiskit2024}
A.~Javadi-Abhari, M.~Treinish, K.~Krsulich, C.~J. Wood, J.~Lishman, J.~Gacon,
  S.~Martiel, P.~D. Nation, L.~S. Bishop, A.~W. Cross, B.~R. Johnson, and J.~M.
  Gambetta, arXiv:2405.08810 (2024).

\bibitem{golden_2022}
J.~Golden, A.~B{\"a}rtschi, D.~O'Malley, and S.~Eidenbenz, ACM Trans. Quantum
  Comput. {\bfseries 3},  8:1 (2022).

\end{thebibliography}

\end{document}